# Reference Publication Year Spectroscopy (RPYS) in practice: A software tutorial

Robin Haunschild* and Lutz Bornmann*$

* Max Planck Institute for Solid State Research

Heisenbergstraße 1,

70569 Stuttgart, Germany.

Email: R.Haunschild@fkf.mpg.de, L.Bornmann@fkf.mpg.de

$ Administrative Headquarters of the Max Planck Society

Science Policy and Strategy Department

Hofgartenstr. 8,

80539 Munich, Germany.

Email: bornmann@gv.mpg.de

Abstract

In course of the organization of Workshop III entitled "Cited References Analysis Using CRExplorer" at the International Conference of the International Society for Scientometrics and Informetrics (ISSI2021), we have prepared three reference publication year spectroscopy (RPYS) analyses: (i) papers published in *Journal of Informetrics*; (ii) papers regarding the topic altmetrics; and (iii) papers published by Ludo Waltman (we selected this researcher since he received the Derek de Solla Price Memorial Medal during the ISSI2021 conference). The first RPYS analysis has been presented live at the workshop and the second and third RPYS analyses have been left to the participants for undertaking after the workshop. Here, we present the results for all three RPYS analyses. The three analyses have shown quite different seminal papers with a few overlaps. Many of the foundational papers in the field of scientometrics (e.g., distributions of publications and citations, citation network and co-citation analyses, and citation analysis with the aim of impact measurement and research evaluation) were retrieved as seminal papers of the papers published in *Journal of Informetrics*. Mainly papers with discussions of the deficiencies of citation-based impact measurements and comparisons between altmetrics and citations were retrieved as seminal papers of the topic altmetrics. The RPYS analysis of the paper set published by Ludo Waltman mainly retrieved papers about network analyses, citation relations, and citation impact measurement.

# Introduction

We organized Workshop III entitled "Cited References Analysis Using CRExplorer" at the International Conference of the International Society for Scientometrics and Informetrics (ISSI2021). We reported about the detailed procedure of this workshop earlier (Haunschild & Bornmann, 2021b). A cited references analysis of the papers published in *Journal of Informetrics* was presented in an interactive part of this workshop. We could not present two other cited references analyses due to time constraints: One on the topic altmetrics, short for alternative metrics such as Twitter or Mendeley counts (Sugimoto, Work, Larivière, & Haustein, 2017), and one on the papers published by Ludo Waltman. The purpose of this contribution is to put the cited references analysis that was performed during the workshop into writing and present the other two applications of cited references analyses. Furthermore, we intend to present these three applications as examples of current good practice of cited references analyses.

Times cited analyses provide a forward perspective from a set of focal papers whereas cited references analyses turn this around and provide a backwards perspective. The impact of publications on specific publication sets (e.g., journals, topics, or oeuvres of researchers) can be analyzed using the cited references' perspective. The focus of citation impact measurement within the cited references' perspective lies on the selected publication set and not the whole literature database (such as Web of Science, that can be cited in principle). This enables researchers to answer questions, such as: Which publications are seminal papers or historical roots of the analyzed publication set? Bornmann and Marx (2013) proposed thus a specific new form of cited references analysis. This new form was named reference publication year spectroscopy (RPYS, Marx, Bornmann, Barth, & Leydesdorff, 2014). The most frequent applications of RPYS include the discovery of seminal papers and historical roots of papers published in a journal, in a topic, or by a researcher.

RPYS analyses consist of different steps: In the first step, the publication set under study has to be collected. In the second step, the publication set with cited references is analyzed with regard to the number of cited references in each reference publication year (RPY). In the third and final step, the RPYS results are analyzed (and interpreted). A plot of the RPYs against the number of cited references shows early peaks where historical roots can be found. Different metrics (e.g., in how many citing years were certain references cited very frequently) can be used to further support the analysis of the RPYS results. The program CRExplorer (see [www.crexplorer.net](www.crexplorer.net)) was introduced by Thor, Marx, Leydesdorff, and Bornmann (2016a) for simplifying and supporting the latter two stages. Advanced indicators that provide new cited references analysis opportunities were included in the capabilities of CRExplorer later (Thor, Bornmann, Marx, & Mutz, 2018).

During our workshop, Rüdiger Mutz and Peter Kokol presented RPYS analyses as invited talks. Their analyses can be found in the literature (Barth, Marx, Bornmann, & Mutz, 2014; Kokol, Zavrsnik, & Blazun Vosner, 2021). In this contribution, we present three RPYS analyses: (1) papers published in *Journal of Informetrics*, (2) papers regarding the topic altmetrics, and (3) papers published by Ludo Waltman (we selected this researcher since he received the Derek de Solla Price Memorial Medal during the ISSI2021 conference). The first one has been presented live at the workshop and data sets for the second and third one have been provided to the participants of the workshop. We are happy to share these data sets also with interested readers on a personal basis. These three RPYS analyses represent examples of the frequent uses of RPYS: Studying the historical roots and seminal papers of a journal, a research field, or a researcher. We encourage the readers to analyze the data sets theirselves to better understand the RPYS procedure.

## Data and methods

### Data

We use publication metadata from Web of Science (WoS; Birkle, Pendlebury, Schnell, & Adams, 2020) by Clarivate Analytics. The data were downloaded in June 2021. We searched for all papers authored by Ludo Waltman, all papers with the topic (title, abstract, and keywords) altmetrics, and all papers published in *Journal of Informetrics*. Among the three data sets, the data set regarding the topic altmetrics should be considered most incomplete because not all publications on the topic altmetrics use this term in their title, abstract, or keywords. However, RPYS analyses are rather robust in this respect. A more complete publication set that was retrieved by a more complicated search query should not yield very different results (Haunschild, Marx, Thor, & Bornmann, 2019).

### Methods

We used CRExplorer (https://crexplorer.net; Thor, Bornmann, Marx, et al., 2018; Thor, et al., 2016a; Thor, Marx, Leydesdorff, & Bornmann, 2016b) for RPYS analysis. CRExplorer can be used interactively and via the script interface. The script interface has the advantage of enabling the reproducibility of the analysis. The following CRExplorer scripts in Figure 1, Figure 2, and Figure 3 can be used to reproduce our results after downloading the corresponding publication data.

```
1  importFile(files: ["Journal_of_Informetrics_pt1.txt", "Journal_of_Informetrics_pt2.txt",
2  "Journal_of_Informetrics_pt3.txt"], type: "WOS", RPY: [1000, 2021, true], PY: [1900, 2021, true], maxCR: 0)
3  cluster(threshold: 0.75, volume: true, page: true, DOI: false)
4  merge()
5  removeCR( N_CR: [0, 9])
6  exportFile(file: "Journal_of_Informetrics_rpys_mincr_10_script_CR.csv", type: "CSV_CR")
7  exportFile(file: "Journal_of_Informetrics_rpys_mincr_10_script_GRAPH.csv", type: "CSV_GRAPH")
8  saveFile(file: "Journal_of_Informetrics_rpys_mincr_10_script.cre")
```

Figure 1: CRExplorer script for RPYS analysis of the papers published in *Journal of Informetrics*

```
1  importFile(files: ["topic_search_altmetrics_pt1.txt", "topic_search_altmetrics_pt2.txt"],
2  type: "WOS", RPY: [1000, 2021, true], PY: [1900, 2021, true], maxCR: 0)
3  cluster(threshold: 0.75, volume: true, page: true, DOI: false)
4  merge()
5  removeCR( N_CR: [0, 4])
6  exportFile(file: "topic_search_altmetrics_rpys_mincr_5_script_CR.csv", type: "CSV_CR")
7  exportFile(file: "topic_search_altmetrics_rpys_mincr_5_script_GRAPH.csv", type: "CSV_GRAPH")
8  saveFile(file: "topic_search_altmetrics_rpys_mincr_5_script.cre")
```

Figure 2: CRExplorer script for RPYS analysis of the papers regarding the topic altmetrics

```
1  importFile(file: "Ludo_Waltman.txt", type: "WOS", RPY: [1000, 2021, true],
2  PY: [1900, 2021, true], maxCR: 0)
3  cluster(threshold: 0.75, volume: true, page: true, DOI: false)
4  merge()
5  removeCR( N_CR: [0, 1])
6  exportFile(file: "Ludo_Waltman_rpys_mincr_2_script_CR.csv", type: "CSV_CR")
7  exportFile(file: "Ludo_Waltman_rpys_mincr_2_script_GRAPH.csv", type: "CSV_GRAPH")
8  saveFile(file: "Ludo_Waltman_rpys_mincr_2_script.cre")
```

Figure 3: CRExplorer script for RPYS analysis of the papers published by Ludo Waltman

The first two lines of the CRExplorer scripts in Figure 1, Figure 2, and Figure 3 import the WoS data. The third line invokes an automatic clustering algorithm. Since cited references data are concerned by variants of the same reference, unifying of variants is an important step in RPYS. We used the clustering algorithm with a Levenshtein threshold of 0.75 accounting for volume and page number. We have had good experience with a Levenshtein threshold of 0.75 in many previous studies: It is a good compromise between clustering references that are different and missing too many references that are the same. DOIs are not used in the clustering of the cited references, since we made the experience that the focus on volume and page number leads to good clustering results (in most of the cases). When books are important in a field or topic, neglecting the page number or manual clustering might be necessary. In the fourth line, the equivalent cited references are merged. The fifth line removes rarely occurring cited references.

Removing of rarely occurring references leads to more pronounced RPYS results. We used different removal thresholds for the three data sets due to the different total number of cited references: (1) All cited references occurring less than ten times were removed from the RPYS analysis of the papers published in *Journal of Informetrics*; (2) all cited references occurring less than five times were removed from the RPYS analysis of the papers regarding the topic altmetrics, and (3) all cited references occurring only once were removed from the RPYS analysis of the papers published by Ludo Waltman. The choice of these thresholds roughly depend on the total number of cited references within the data sets: The higher the number of cited references, the higher the appropriate threshold. When inspecting an unknown data set, one should start by removing the cited references that occur only once. This threshold should be raised until a peak structure is visible in the spectrogram. Early peaks might disappear if too high threshold values are used.

Lines six to eight in the scripts export the results: A CSV file with the remaining cited references is exported in line six. A CSV file with the number of cited references per RPY for plotting the RPYS spectrogram is exported in line seven. A CRE file that can be opened later in CRExplorer is saved in line eight. It is also possible to make multiple exports of different data sets, e.g., the user can export the CSV files with all cited references included. Afterwards remove the singly cited references and export this data set, and subsequently remove all cited references that were cited less than five times and export that data set and so on. The exported CSV files can be inspected in the interactive CRExplorer and the most appropriate result can be analyzed further.

Table 1 shows relevant data regarding cited references and citing publications of the three RPYS analyses. The CSV file that is written in line 7 can be imported in other programs for producing high-quality and customized graphics.

Table 1: Relevant data regarding cited references and citing publications of the three RPYS analyses

|  | *Journal of Informetrics* | Altmetrics | Ludo Waltman |
|---|---:|---:|---:|
| Total number of non-distinct cited references | 41,878 | 32,411 | 2,737 |
| Minimum RPY | 1230 | 1665 | 1896 |
| Maximum RPY | 2021 | 2021 | 2021 |
| Number of citing publications | 1,106 | 896 | 101 |
| Minimum citing year | 2007 | 2012 | 2005 |
| Maximum citing year | 2021 | 2021 | 2021 |
| Total number of distinct cited references before merging | 21,490 | 17,621 | 1,710 |
| Number of different RPYs | 128 | 107 | 68 |
| Number of different citing years | 15 | 10 | 17 |
| Total number of distinct cited references after merging | 21,443 | 17,587 | 1,710 |

| Total number of distinct cited references after removal of rarely occurring cited references | | 504 | 826 | 471 |

We analyzed the RPYS spectrograms regarding relevant peaks of the five-year median deviation. Tukey's fences (Tukey, 1977) were used to support the identification of the most important peaks: Important peaks are flagged based on the interquartile range of the median deviations (Thor, Bornmann, & Haunschild, 2018). The relevant cited references under the peaks can be determined by the N_CR values (number of cited references) and PERC_YR values (proportion of the number of cited references within this RPY, i.e., N_CR/NCR). However, proper selection of the relevant cited references requires some experience. We hope that the selection we made for the three example RPYS analyses provides helpful guidelines for the readers: (i) Cited references with rather high PERC_YR values should be selected because they contribute a significant proportion of the entire peak. (ii) Multiple top cited references of peaks should be selected if they have rather similar N_CR values. If a larger gap occurs in the N_CR values within the peak. cited references above the gap should be selected.

We used the N_TOP10 indicator to discover highly influential cited references of the three publication sets. The advanced indicator N_TOP10 (Thor, Bornmann, Marx, et al., 2018) counts the number of citing years in which the corresponding cited reference has been cited very frequently (i.e., within the top-10%). We expect that highly influential cited references have values of at least half of the maximal N_TOP10 value, i.e., more than half of the number of different citing years in Table *1*. Furthermore, we checked the cited references with the highest

N_CR values. Other advanced indicators, for example N_TOP1 (i.e., within the top-1%) and N_TOP0_1 (i.e., within the top-0.1%), are available if needed (Thor, Bornmann, Haunschild, & Leydesdorff, 2020).

The RPYS spectrograms were plotted using R (R Core Team, 2018) with the R package 'BibPlots' (Haunschild, 2021). As part of the preparation and post-processing of our workshop, the R package 'BibPlots' was extended. The new function 'rpys_bl' was added for producing RPYS bar-line charts. Such RPYS charts display the number of cited references as a bar-plot with a line-plot overlay for the median deviation. This is the plot style used in the RPYS spectrograms in this paper. In addition to the static plots in this paper, we produced interactive RPYS graphs using the R package 'dygraphs' (Vanderkam, Allaire, Owen, Gromer, & Thieurmel, 2018).

## Results

In the following three subsections, we present the results of the three RPYS analyses prepared for the workshop. The first one has been performed during the workshop, and the other ones were left to the participants to try on their own afterwards.

### RPYS analysis of the papers published in *Journal of Informetrics*

Figure 4 presents the spectrogram from the RPYS analysis of the papers published in *Journal of Informetrics*. The first reference cited at least ten times is Hazen (1914). This cited reference is cited in bibliometrics usually when Hazen percentiles are used, e.g., for citation impact normalization purposes (see for example Bornmann & Haunschild, 2018). The most important peaks in Figure 4 are marked with a red star, and the RPY is written next to it. An interactive version of Figure 4 is available at: https://s.gwdg.de/EmA74A. Readers can zoom into specific regions of the RPYS spectrogram using the interactive version. Crosshairs are located at the data points closest to the mouse cursor for improved usability.

When the mouse cursor hovers over the tooltip flag on top of the bars, a tooltip appears that shows up to five most referenced cited references of that RPY with additional advanced indicators: PERC_YR, N_TOP10, and N_PYEARS (number of different citing years in which this cited reference has been cited). The most important cited references from the marked RPYs are listed in Table 2.

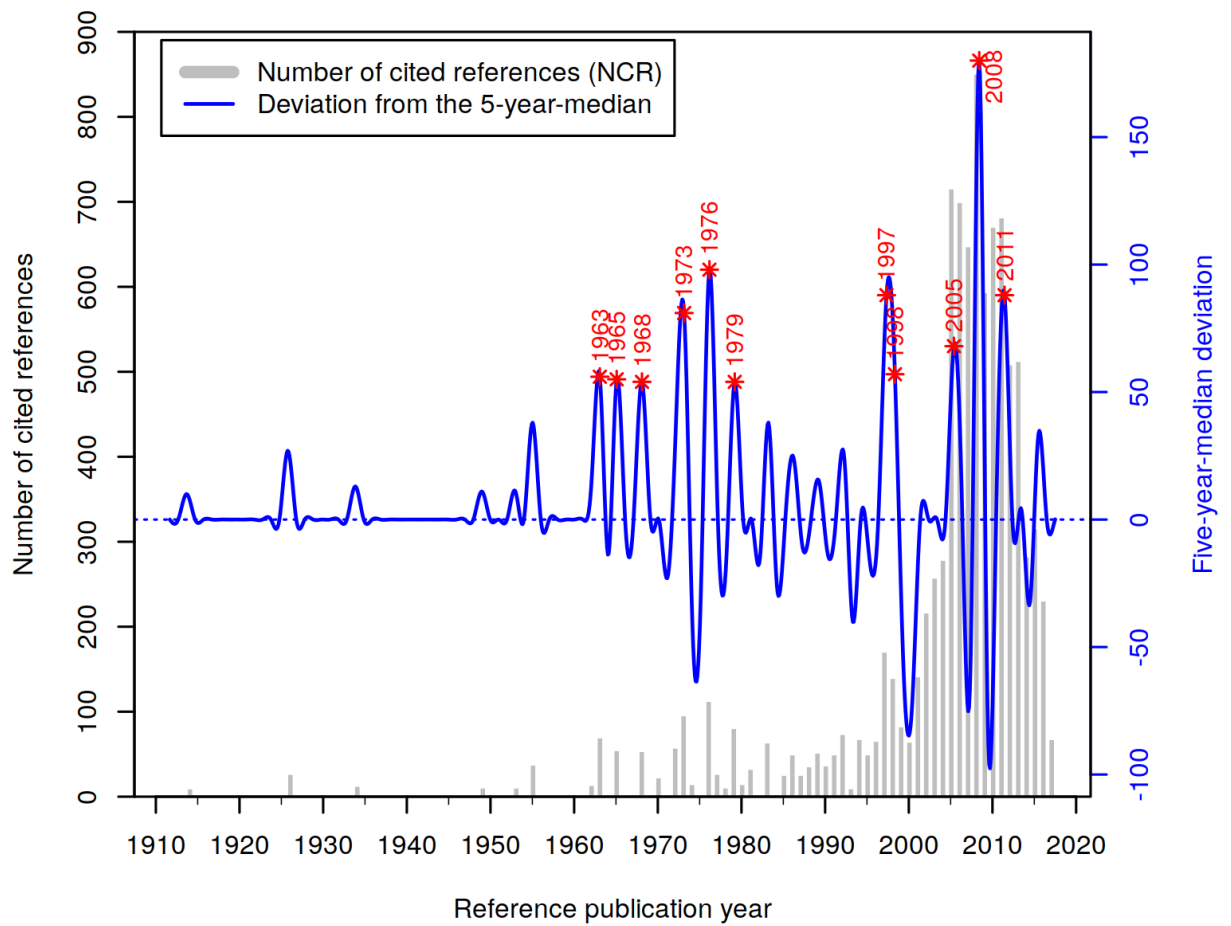

Figure 4: Spectrogram from the RPYS analysis of the papers published in *Journal of Informetrics*. An interactive version is available at: https://s.gwdg.de/EmA74A.

Table 2: Relevant cited references that correspond to the peaks marked in Figure 4 with RPY and N_CR ordered primarily by RPY and secondarily by N_CR

| Nº | Cited Reference | RPY | N_CR |
|---|---|---|---|
| CR1 | Kessler, M. M., 1963, Am. Doc., V14, P10, DOI 10.1002/asi.5090140103. Bibliographic Coupling Between Scientific Papers. | 1963 | 30 |
| CR2 | de Solla Price, D. J., 1963, Little Science Big Science | 1963 | 28 |
| CR3 | de Solla Price, D. J., 1965, Science, V149, P510. Networks of Scientific Papers. | 1965 | 55 |
| CR4 | Merton, R. K., 1968, Science, V159, P56, DOI 10.1126/science.159.3810.56. The Matthew Effect in Science: The Reward and Communication Systems of Science are Considered. | 1968 | 54 |
| CR5 | Small, H., 1973, J. Am. Soc. Inform. Sci., V24, P265, DOI 10.1002/asi.4630240406. Co-Citation in the Scientific Literature: A New Measure of the Relationship Between Two Documents. | 1973 | 38 |
| CR6 | Merton, R. K., 1973, The sociology of science: Theoretical and empirical investigations | 1973 | 29 |
| CR7 | Pinski, G., 1976, Inform. Process. Manag., V12, P297, DOI 10.1016/0306-4573(76)90048-0. Citation Influence for Journal Aggregates of Scientific Publications: Theory, with Application to the Literature of Physics. | 1976 | 50 |
| CR8 | de Solla Price, D. J., 1976, J. Am. Soc. Inform. Sci., V27, P292, DOI 10.1002/asi.4630270505. A General Theory of Bibliometric and Other Cumulative Advantage Processes. | 1976 | 49 |
| CR9 | Garfield, E., 1979, Scientometrics, V1, P359, DOI 10.1007/BF02019306. Is Citation Analysis a Legitimate Evaluation Tool? | 1979 | 32 |
| CR10 | Freeman, L. C., 1979, Soc. Networks, V1, P215, DOI 10.1016/0378-8733(78)90021-7. Centrality in Social Networks Conceptual Clarification. | 1979 | 30 |
| CR11 | Seglen, P. O., 1997, Brit. Med. J., V314, P498. Why the Impact Factor of Journals Should not be Used for Evaluating Research. | 1997 | 51 |
| CR12 | Katz, J. S., 1997, Res. Policy, V26, P1, DOI 10.1016/S0048-7333(96)00917-1. What is Research Collaboration? | 1997 | 47 |
| CR13 | Brin, S., 1998, Comp. Networks ISDN Syst., V30, P107, DOI 10.1016/S0169-7552(98)00110-X. The Anatomy of a Large-Scale Hypertextual Web Search Engine. | 1998 | 40 |
| CR14 | Redner, S., 1998, Eur. Phys. J. B, V4, P131, DOI 10.1007/s100510050359. How Popular is Your Paper? An Empirical Study of the Citation Distribution. | 1998 | 28 |
| CR15 | Hirsch, J. E., 2005, PNAS USA, V102, P16569, DOI 10.1073/pnas.0507655102. An Index to Quantify an Individual's Scientific Research Output. | 2005 | 262 |
| CR16 | Radicchi, F., 2008, PNAS USA, V105, P17268, DOI 10.1073/pnas.0806977105. Universality of Citation Distributions: Toward an Objective Measure of Scientific Impact. | 2008 | 81 |
| CR17 | Bornmann, L., 2008, J. Doc., V64, P45, DOI 10.1108/00220410810844150. What do Citation Counts Measure? A Review of Studies on Citing Behavior. | 2008 | 61 |
| CR18 | Waltman, L., 2011, J. Informetr., V5, P37, DOI 10.1016/j.joi.2010.08.001. Towards a New Crown Indicator: Some Theoretical Considerations. | 2011 | 71 |

All publications in Table 2 are well-known landmark publications in scientometrics. In CR1, bibliographic coupling has been proposed as a method for grouping publications by M. M. Kessler. CR2 is a classical book of collected lectures given by D. J. de Solla Price entitled "Little Science, Big Science". This lecture collection discusses science and its place in society. CR3 is the work of D. J. de Solla Price regarding networks of scientific publications and about incidences of references and citations. CR4 is R. K. Merton's proposal of the Matthew effect of accumulated advantage in science. CR5 is H. Small's proposal to use co-citations as a measure of relationships between scientific publications. CR6 is R. K. Merton's book "The sociology of science: Theoretical and empirical investigations" explaining many well-known concepts such as the Matthew effect.

CR7 is the contribution by G. Pinski and F. Narin regarding citation influence of scientific publications on the journal basis. This work constitutes the basis for Google's PageRank algorithm (Page, 2001). In CR8, D. J. de Solla Price presented the cumulative advantage distribution as an underlying theory for Bradford's law, Lotka's law, and Zipf's law. CR9 is E. Garfield's discussion on the use of citation analysis to rate scientific performance. CR 10 is L. C. Freeman's review on measures of structural centrality in social networks. In CR11, P. O. Seglen argued against the use of the journal impact factor in research evaluation. In CR12, J. S. Katz and B. R. Martin discussed research collaboration on a conceptual level and point out that co-authorship is not necessarily adequate to describe collaboration. In CR13, S. Brin and L. Page describe a prototype of the Google search engine.

CR14 is S. Redner's empirical analysis of the citation distribution of papers published in 1981 and indexed by the Institute for Scientific Information (one of the predecessors of Clarivate Analytics). In CR15, J. E. Hirsch proposed the popular although heavily criticized h index for measuring the performance of individual researchers. Based on an analysis of common citation distributions, F. Radicchi proposed a generalized h index in CR 16. CR17 is L. Bornmann's and H.-D. Daniel's discussion on the question what citation counts are measuring based on their review of studies on citation habits. In CR18, L. Waltman et al. proposed a new crown indicator (mean normalized citation score, MNCS) and presented an analysis of its formal mathematical properties.

Two cited references in Table 2 have the highest possible N_TOP10 value of 15 in our data set. One of them is CR15 (the introduction of the h index), and the other one is a theoretical analysis by L. Egghe of the g index, a modified version of the h index (Egghe, 2006), with N_CR=109. Thus, Egghe (2006) received less than half the number of citations compared with CR15. Therefore, the peak in 2005 and CR15 itself dominate the neighboring years. It is mainly a matter of choice how detailed RPYS analyses are performed. A more in-depth analysis of the data set might have listed Egghe (2006) in Table 2. However, our aim with this contribution is more to provide guidance how RPYS analyses should be performed than to analyze the example data sets in high detail. The other cited references that surpassed half of the maximal possible N_TOP10 value are also listed in Table 2 (CR16, CR17, and CR18) except for H. F. Moed's proposal of a new journal impact metric, the source normalized impact per paper, SNIP (Moed, 2010). Only two cited references reached N_CR values above 100: J. Hirsch's paper with N_CR=262 (CR15) and Egghe (2006) with N_CR=109 (not listed in Table 2).

This example shows the importance of looking also into regions of the RPYS spectrogram with high N_CR values even if the median deviation is low. Usually, as in this case, such highly influential cited references are found when sorting the list of cited references by N_CR or N_TOP10 values. We recommend this approach in addition to only looking at extreme peaks. Alternatively, one could inspect each RPY with a sizable N_CR value separately. If many additional cited references are found by these approaches, separate tables should be provided for presenting them. In our case here, only two additional cited references were found by the analyses. Thus, we recommend mentioning those cited references in the text and formally citing them.

RPYS analysis of the papers regarding the topic altmetrics

The spectrogram that results from our RPYS analysis of the papers published regarding the topic altmetrics is shown in Figure 5. The first reference cited at least five times is Lotka (1926). Based on his analysis of the productivity of chemists and physicists, A. Lotka found that statistically $1/n^2$ scientists are making $n$ contributions to science, i.e., many scientists publish few papers and few scientists publish many papers. This became known as Lotka's law, see also CR8 in Table 2. Red stars mark the most important peaks in Figure 5. Additionally, the corresponding RPY is written next to the star. Table 3 lists the most important cited references from the marked RPYs in Figure 5.

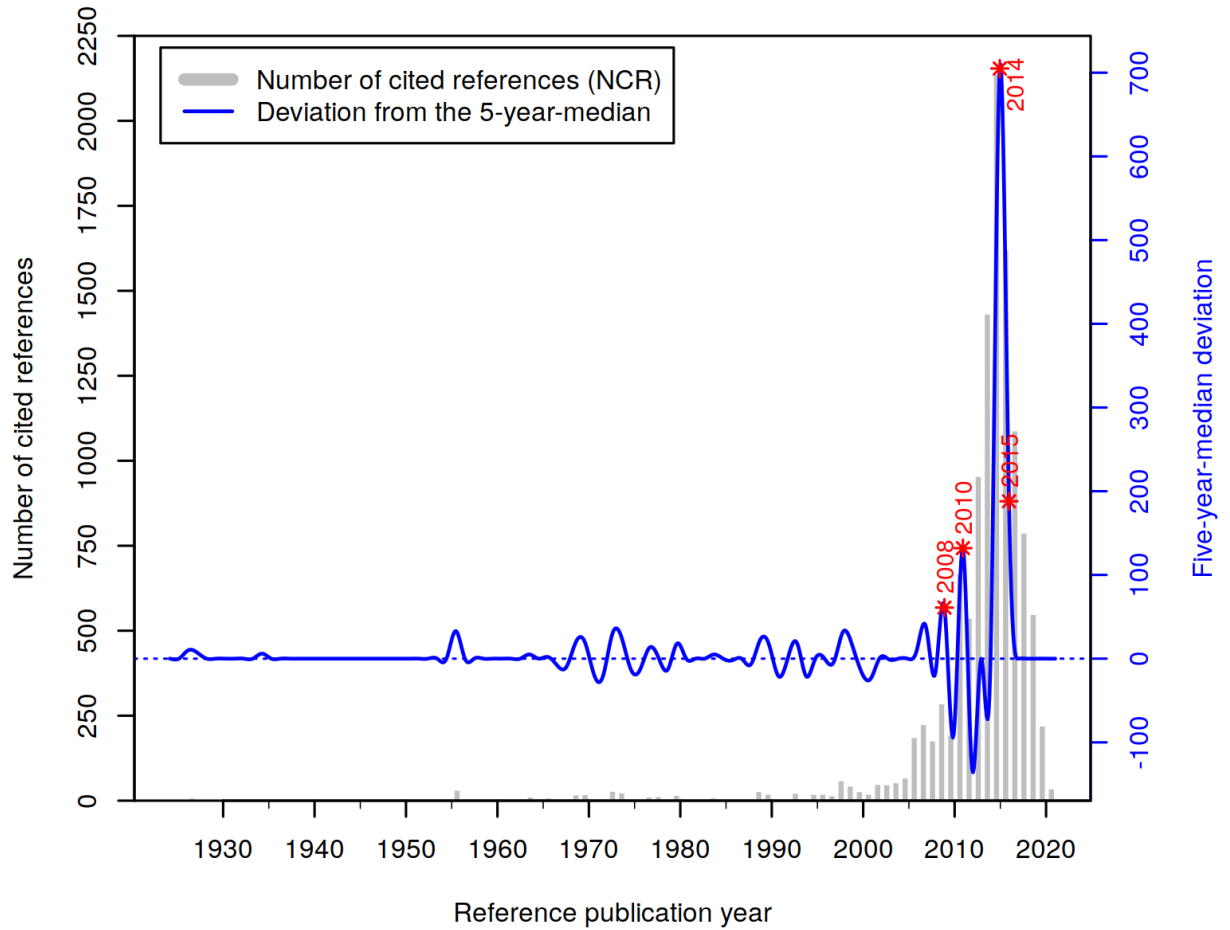

Figure 5: Spectrogram from the RPYS analysis of the papers published regarding the topic altmetrics. An interactive version is available at: https://s.gwdg.de/zaPCOY.

Table 3: Relevant cited references that correspond to the peaks marked in Figure 5 with RPY and N_CR ordered primarily by RPY ascending and secondarily by N_CR descending

| Nº | Cited Reference | RPY | N_CR |
|---|---|---|---|
| CR19 | Bornmann, L., 2008, J. Doc., V64, P45, DOI 10.1108/00220410810844150. What do Citation Counts Measure? A Review of Studies on Citing Behavior. | 2008 | 36 |
| CR20 | Priem, J., 2010, Altmetrics Manifesto | 2010 | 220 |
| CR21 | Bornmann, L., 2014, J. Informetr., V8, P895, DOI 10.1016/j.joi.2014.09.005. Do Altmetrics Point to the Broader Impact of Research? An Overview of Benefits and Disadvantages of Altmetrics. | 2014 | 145 |
| CR22 | Zahedi, Z., 2014, Scientometrics, V101, P1491, DOI 10.1007/s11192-014-1264-0. How Well Developed are Altmetrics? A Cross- | 2014 | 141 |

| | Disciplinary Analysis of the Presence of 'Alternative Metrics' in Scientific Publications. | | |
|---|---|---|---|
| CR23 | Costas, R., 2015, J. Assoc. Inf. Sci. Tech., V66, P2003, DOI 10.1002/asi.23309. Do "Altmetrics" Correlate with Citations? Extensive Comparison of Altmetric Indicators with Citations from a Multidisciplinary Perspective. | 2015 | 201 |

CR19 in Table 3 is actually the same cited reference as CR 17 in Table 2 but we provide a new CR number for this cited reference because the N_CR values in both RPYS analyses differ due to the different data set. CR20 shows one of the main strengths of cited references analyses. CR20 is the altmetrics manifesto by J. Priem, D. Taraborelli, P. Groth, C. Neylon that provided early reasons for the need of additional new metrics. This cited reference is a webpage, i.e., a non-source reference of citation databases like WoS. A topic search alone would not yield this digital document. The RPYS analysis is able to discover such important non-source references. In fact, CR20 is the most cited reference in Table 3. CR21 is a discussion about the relationship between altmetrics and broader impact (e.g., societal impact) and an overview of advantages and downsides of altmetrics by L. Bornmann. CR22 is a status report about the development of altmetrics by Z. Zahedi, R. Costas, and P. Wouters. They analyzed the presence of altmetrics across disciplines and calculated correlations between altmetrics and citations. CR 23 is a follow-up study of CR22 that uses a larger data set by the same authors in a slightly different order.

There are 22 cited references that surpass the N_TOP10 threshold for this analysis. Only four (CR20, CR21, CR22, and CR23) of these 22 cited references appeared in Table 3 and were just discussed. In the previous section where we discussed the RPYS analysis of the papers published in the *Journal of Informetrics*, only two additional cited references were found by analyzing the N_TOP10 values. In situations with many additional cited references like the current one, we recommend listing these additional cited references in another table. Thus, the additional cited references are listed in Table 4.

Table 4: Cited references with N_TOP10 values higher than five (half of the maximal possible value) ordered descending primarily by N_TOP10 and secondarily by N_CR

| No | Cited Reference | RPY | N_CR | N_TOP10 |
|---|---|---|---|---|
| CR24 | Eysenbach, G., 2011, J. Med. Internet Res., V13, DOI 10.2196/jmir.2012. Can Tweets Predict Citations? Metrics of Social Impact Based on Twitter and Correlation with Traditional Metrics of Scientific Impact. | 2011 | 170 | 9 |
| CR25 | Thelwall, M., 2013, PLoS One, V8, DOI 10.1371/journal.pone.0064841. Do Altmetrics Work? Twitter and Ten Other Social Web Services. | 2013 | 238 | 8 |
| CR20 | Priem, J., 2010, Altmetrics Manifesto | 2010 | 220 | 8 |
| CR26 | Haustein, S., 2014, J. Assoc. Inf. Sci. Tech., V65, P656, DOI 10.1002/asi.23101. Tweeting Biomedicine: An Analysis of Tweets and Citations in the Biomedical Literature. | 2014 | 126 | 8 |
| CR27 | Li, X. M., 2012, Scientometrics, V91, P461, DOI 10.1007/s11192-011-0580-x. A Correlation Comparison Between Altmetric Attention Scores and Citations for Six PLOS Journals. | 2012 | 109 | 8 |
| CR28 | Neylon, C., 2009, PLoS Biol., V7, DOI 10.1371/journal.pbio.1000242. Article-Level Metrics and the Evolution of Scientific Impact. | 2009 | 41 | 8 |
| CR23 | Costas, R., 2015, J. Assoc. Inf. Sci. Tech., V66, P2003, DOI 10.1002/asi.23309. Do "Altmetrics" Correlate with Citations? Extensive Comparison of Altmetric Indicators with Citations from a Multidisciplinary Perspective. | 2015 | 201 | 7 |
| CR21 | Bornmann, L., 2014, J. Informetr., V8, P895, DOI 10.1016/j.joi.2014.09.005. Do Altmetrics Point to the Broader Impact of Research? An Overview of Benefits and Disadvantages of Altmetrics. | 2014 | 145 | 7 |
| CR22 | Zahedi, Z., 2014, Scientometrics, V101, P1491, DOI 10.1007/s11192-014-1264-0. How Well Developed are Altmetrics? A Cross-Disciplinary Analysis of the Presence of 'Alternative Metrics' in Scientific Publications. | 2014 | 141 | 7 |
| CR29 | Mohammadi, E., 2014, J. Assoc. Inf. Sci. Tech., V65, P1627, DOI 10.1002/asi.23071. Mendeley Readership Altmetrics for the Social Sciences and Humanities: Research Evaluation and Knowledge Flows. | 2014 | 115 | 7 |
| CR30 | Haustein, S., 2014, Scientometrics, V101, P1145, DOI 10.1007/s11192-013-1221-3. Coverage and Adoption of Altmetrics Sources in the Bibliometric Community. | 2014 | 93 | 7 |
| CR31 | Hirsch, J. E., 2005, PNAS USA, V102, P16569, DOI 10.1073/pnas.0507655102. An Index to Quantify an Individual's Scientific Research Output. | 2005 | 72 | 7 |

| CR32 | Piwowar, H., 2013, Nature, V493, P159, DOI 10.1038/493159a. Value All Research Products. | 2013 | 71 | 7 |
| --- | --- | --- | --- | --- |
| CR33 | Shuai, X., 2012, PLoS One, V7, DOI 10.1371/journal.pone.0047523. How the Scientific Community Reacts to Newly Submitted Preprints: Article Downloads, Twitter Mentions, and Citations. | 2012 | 67 | 7 |
| CR34 | Priem, J., 2012, PLoS One, V7, DOI 10.1371/journal.pone.0048753. The Altmetrics Collection. | 2012 | 62 | 7 |
| CR35 | Priem, J.., 2010, 1st Monday, V15. Scientometrics 2.0: New Metrics of Scholarly Impact on the Social Web. | 2010 | 59 | 7 |
| CR36 | Wouters, P., 2012, Users, Narcissism and Control—Tracking the Impact of Scholarly Publications in the 21 st Century | 2012 | 59 | 7 |
| CR37 | Bornmann, L., 2008, J. Doc., V64, P45, DOI 10.1108/00220410810844150. What Do Citation Counts Measure? A Review of Studies on Citing Behavior. | 2008 | 36 | 7 |
| CR38 | Sud, P., 2014, Scientometrics, V98, P1131, DOI 10.1007/s11192-013-1117-2. Evaluating Altmetrics. | 2014 | 95 | 6 |
| CR39 | Haustein, S., 2015, PLoS One, V10, DOI 10.1371/journal.pone.0120495. Characterizing Social Media Metrics of Scholarly Papers: The Effect of Document Properties and Collaboration Patterns. | 2015 | 95 | 6 |
| CR40 | Priem, J., 2012, Altmetrics in the wild: Using social media to explore scholarly impact | 2012 | 48 | 6 |
| CR41 | Haustein, S., 2011, J. Informetr., V5, P446, DOI 10.1016/j.joi.2011.04.002. Applying Social Bookmarking Data to Evaluate Journal Usage. | 2011 | 44 | 6 |

In CR24, G. Eysenbach studied the suitability of Twitter counts as an early proxy for citation counts. In CR25, M. Thelwall et al. studied the correlations between eleven altmetrics and citations. Actually, CR25 is the most-frequently referenced cited reference in the altmetrics literature. CR26 presents an analysis by S. Haustein et al. of the relationship between tweets and citations in the biomedical literature. CR27 by Li et al. checked the validity of using bookmark counts in online reference managers for research evaluation. In CR28, C. Neylon and S. Wu discussed the need of article-level metrics that accumulate faster than citations thereby paving the way towards altmetrics. In CR29, E. Mohammadi and M. Thelwall compared the number of times papers from different disciplines in Social Sciences and Humanities have been bookmarked in the online reference manager platform Mendeley with the numbers of times the same papers have been cited. In CR30, S. Haustein et al. analyzed the altmetrics activity regarding the papers authored by 57 of the presenters at the 2010 STI conference in Leiden, The Netherlands. CR31 in Table 4 is actually the same cited reference as CR15 in Table 2, but we assigned a different CR number to this cited reference here because the N_CR value is different within the topic of altmetrics from that in the papers published in *Journal of Informetrics*.

H. Piwowar discussed the implications of a small change of words (applicants are asked to list their "research products" rather than their "publications") in the US National Science Foundation's funding guidelines in CR32 in Table 4. In CR33, X. Shuai, A. Pepe, and J. Bollen studied the altmetrics activity regarding selected arXiv preprints. In CR34, J. Priem, P. Groth, and D. Taraborelli promoted the use of altmetrics for tracking, describing, and analyzing various processes in which scholars are involved (e.g., publishing, teaching, reading, and recommending). CR35 by J. Priem and B. M. Hemminger contains the description of different altmetrics and their possible use before the term altmetrics was coined. In CR36, P. Wouters and R. Costas discussed possible use cases of altmetrics on a broad basis. CR37 in Table 4 is the same cited reference as CR17 in Table *2* just with a different N_CR value due to the different set of citing papers.

CR38 in Table 4 by P. Sud and M. Thelwall contains a discussion about the potential applicability of altmetrics for research evaluation. In CR39, S. Haustein, R. Costas, and V. Larivière presented an analysis of incidence of selected altmetrics with respect to different document characteristics, e.g., document type, number of pages, length of title, and scientific discipline. In CR40, J. Priem, H. Piwowar, and B. M. Hemminger presented an analysis of the frequencies in which papers published by the Public Library of Science (PLoS) were mentioned in selected altmetrics sources. In CR41, S. Haustein and T. Siebenlist proposed to use bookmarking data from online reference managers as a proxy for journal readership. All cited references with N_CR values above 100 are included in Table 4.

## RPYS analysis of the papers published by Ludo Waltman

The RPYS analysis of the papers published by Ludo Waltman yields the spectrogram that is shown in Figure 6. The first cited reference that was cited at least twice is the work of Moskowitz (1958) who analyzed redundancy networks. There is only one extreme peak in Figure 6 that covers two RPYs. Both RPYs are labeled and marked with a red star. There are only three main cited references from these two RPYs, all of them referenced at least ten times. None of the cited references reached the threshold of N_TOP10=9 (derived from 17 citing years). The reason might be that we chose the publication output of a still rather young researcher. Therefore, we present the cited references that were referenced at least ten times in Table 5.

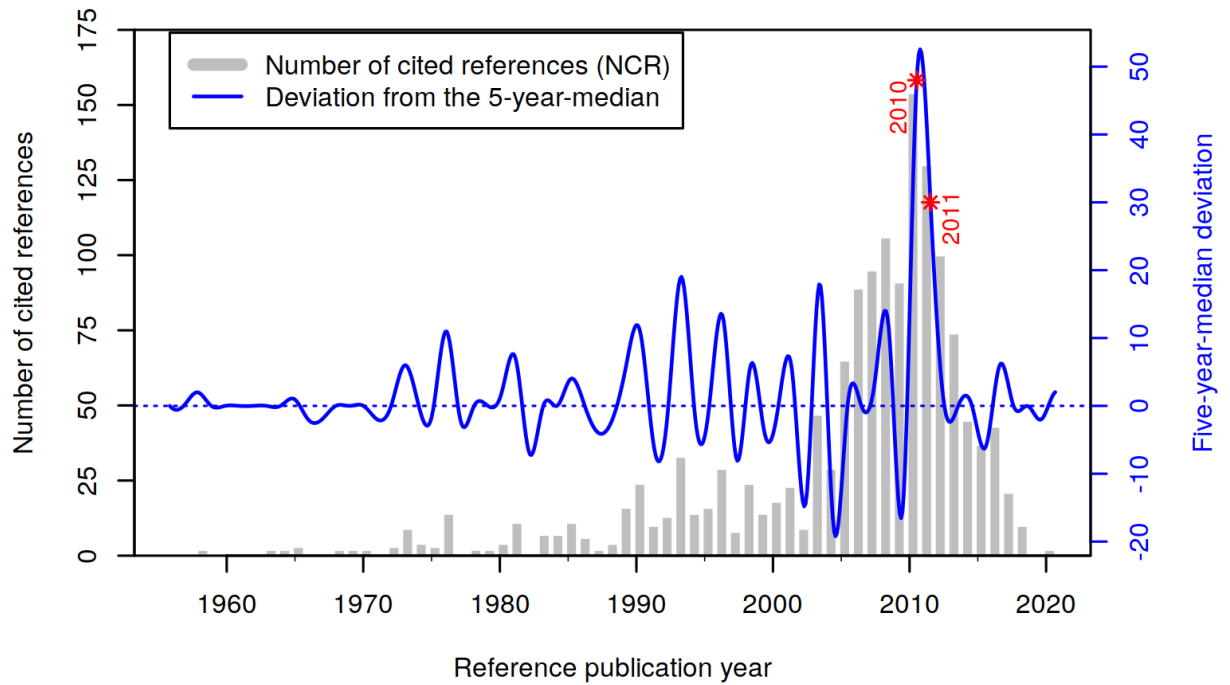

Figure 6: Spectrogram from the RPYS analysis of the papers by Ludo Waltman. An interactive version is available at: https://s.gwdg.de/jdkgRn.

Table 5: Cited references with N_CR values of at least ten from the RPYS analysis of the papers published by Ludo Waltman with RPY, N_CR, and N_TOP10 ordered by N_CR descending

| No | Cited Reference | RPY | N_CR | N_TOP10 |
|---|---|---|---|---|
| CR42 | Waltman, L., 2012, J. Am. Soc. Inf. Sci. Tech., V63, P2378, DOI 10.1002/asi.22748. A New Methodology for Constructing a Publication-Level Classification System of Science. | 2012 | 20 | 7 |
| CR43 | van Eck, N. J., 2010, Scientometrics, V84, P523, DOI 10.1007/s11192-009-0146-3. Software Survey: Vosviewer, a Computer Program for Bibliometric Mapping. | 2010 | 18 | 6 |
| CR44 | Waltman, L., 2011, J. Informetr., V5, P37, DOI 10.1016/j.joi.2010.08.001. Towards a New Crown Indicator: Some Theoretical Considerations. | 2011 | 17 | 5 |
| CR45 | Hirsch, J. E., 2005, PNAS USA, V102, P16569, DOI 10.1073/pnas.0507655102. An Index to Quantify an Individual's Scientific Research Output. | 2005 | 14 | 5 |
| CR46 | van Eck, N. J., 2013, PLoS One, V8, DOI 10.1371/journal.pone.0062395. Citation Analysis May Severely Underestimate the Impact of Clinical Research as Compared to Basic Research. | 2013 | 12 | 4 |
| CR47 | Zitt, M., 2008, J. Am. Soc. Inf. Sci. Tech., V59, P1856, DOI 10.1002/asi.20880. Modifying the Journal Impact Factor by Fractional Citation Weighting: The Audience Factor. | 2008 | 11 | 3 |
| CR48 | Peters, H. P. F., 1993, Res. Policy, V22, P23, DOI 10.1016/0048-7333(93)90031-C. Co-Word-Based Science Maps of Chemical Engineering. Part I: Representations by Direct Multidimensional Scaling. | 1993 | 10 | 3 |
| CR49 | Newman, M. E. J., 2004, Phys. Rev. E, V69, DOI [(I) 10.1103/PhysRevE.69.026113, (II) 10.1103/PhysRevE.69.066133]. (I) Finding and Evaluating Community Structure in Networks. (II) Fast Algorithm for Detecting Community Structure in Networks. | 2004 | 10 | 2 |
| CR50 | Lundberg, J., 2007, J. Informetr., V1, P145, DOI 10.1016/j.joi.2006.09.007. Lifting the Crown—Citation Z-Score. | 2007 | 10 | 3 |
| CR51 | Waltman, L., 2010, J. Informetr., V4, P629, DOI 10.1016/j.joi.2010.07.002. A Unified Approach to Mapping and Clustering of Bibliometric Networks. | 2010 | 10 | 2 |

In CR42 in Table 5, L. Waltman and N. J. van Eck proposed an algorithm for clustering publications on the basis of direct citation relationships. CR43 by N. J. van Eck and L. Waltman is a survey about the visualization software VOSviewer which is very popular in bibliometrics. CR44 and CR45 in Table 5 are actually the same cited references as CR18 and CR15 in Table 2. In CR46, N. J. van Eck et al. reported that citation habits vary considerably in some medical WoS subject categories. They point out that clinical research is cited less frequently than basic or diagnostic research. CR47 by M. Zitt and H. Small is the proposal of the audience factor, a variant of the journal impact factor with fractional citation weighting.

In CR48 in Table 5, H. P. F. Peters and P. Wouters proposed a new approach of term co-occurrence mapping. A manual check has shown that most of the times, both papers that are related to the DOIs of CR49 have been cited together in the studied data set. These cited reference companions presented (i) a fast algorithm for detecting community structures in networks and (ii) a set of algorithms for finding and evaluating community structures in networks. The similar topics of the two cited references, both published in *Physical Review E* in 2004, explain why they were frequently cited together. In CR50, J. Lundberg proposed a paper-based field-normalized logarithmized citation score as an alternative to the so called 'crown indicator'. In CR51, L. Waltman, N. J. van Eck, and E. C. M. Noyons discussed the visualization of similarities (VOS) mapping and clustering technique that was made available to the scientific community within VOSviewer (see CR43).

Discussion and conclusions

In this paper, we have presented three different RPYS analyses that were prepared for and/or presented at the workshop III "Cited References Analysis Using CRExplorer" at the 18[th] ISSI conference: (i) one analysis of the papers published in a journal (*Journal of Informetrics*), (ii) one analysis of a topic (altmetrics), and (iii) one analysis of the papers published by a researcher (Ludo Waltman). Those are three different types of publication sets. These differences are (mainly) responsible for the different approaches we took in the RPYS analyses. The proper approach for an RPYS analysis has to be determined by the user based on the inspection of the data set at hand. We provided the three different examples with the aim to help the reader to find the proper approach for performing their own RPYS analyses. The method RPYS is very versatile; there is no simple rule how a particular data set should be analyzed. However, we hope that this contribution serves as a helpful guidance for other researchers.

The seminal publications of the three publication sets were discussed. The seminal papers of *Journal of Informetrics* can be summarized as many of the foundational papers in the field of scientometrics (e.g., distributions of publications and citations), co-citation and citation network analyses, and citation analysis with the aim of research evaluation and impact measurement. The seminal papers of the topic altmetrics mainly comprise discussions of the deficiencies of citation-based impact measurement and comparisons between altmetrics and citations. The seminal papers of the paper set published by Ludo Waltman are mainly about citation relations, network analyses, and citation impact measurement.

RPYS analyses provide a different perspective than a times cited analysis. One of the main advantages is that also non-source documents can be found if they are included in the indexed list of cited references, like in WoS. Such documents cannot be found using a topic search. In this paper, we demonstrate this advantage based on the altmetrics data set. We expect that this advantage will become more important in the (near) future. Since researchers are increasingly active in writing other texts than classical journal papers or book chapters such as blogs, RPYS is able to reveal the importance of these texts for research in fields, on topics or by researchers. Thus, RPYS is very useful for analyzing the relevance of new text formats for scientific referencing or communication.

One of the advantages of RPYS is that is a very robust method. In our experience, incomplete data sets (i.e., data sets not including all papers on a topic, from a researcher or in a field) yield rather similar results, see for example Haunschild, et al. (2019). A few missing or not relevant publications usually do not distort the general conclusions. However, too many wrong publications in the data set can cause wrong peaks to appear, it is a better strategy to exclude some relevant publications in the data set than to include too many irrelevant ones. However, RPYS is not sensitive to changes in the delineation of the data set.

The main limitation of RPYS is that one is only able to discuss the most-prominent cited references, either those responsible for the main peaks, those with extreme high numbers of cited references, or those with a track record of being highly referenced in many citing years. The latter is very closely related to the method multi-RPYS that "extends the impact of these [RPYS] techniques by first segmenting the data in terms of the publication years of the citing sets, performing the standard RPYS analysis within each set and then rank transforming the de-trended results to compare influential references across the history of the citing set" (Comins & Leydesdorff, 2017, p. 1498).

We invite the readers to use this study as a guide to discover RPYS using the data sets presented here for a more in-depth study of their choice. If readers are more interested in the historical roots and seminal papers of a certain aspect of the topics discussed here, the RPYS data can be inspected with a special focus on such a topic. Successful replication of the analyses presented here can be a good starting point for new RPYS analyses of other topics, journals, and researchers. A good data set for a first RPYS analysis with another publication set would be the publications of one's own research field, since the RPYS results can be well interpreted then.

The selection of peaks is another limitation of RPYS. Usage of Tukey's fences is only one of the possible choices for peak selection. Another possible option is the selection based on visual inspection. However, this is a rather subjective option. The interested reader could browse through the other peaks not discussed here using the interactive versions of the RPYS spectrograms to find other possibly relevant seminal publications.

RPYS analyses should be interpreted against the backdrop of knowledge about the scientific topic of the analyzed data set. Experts can reconcile RPYS results with their knowledge of the analyzed field. RPYS results might also substantiate or refine prior knowledge about the field. RPYS analyses can discover surprising cited references that experts would not have guessed in the first place.

**Acknowledgments**

A preprint of this paper that has been used in the ISSI workshop has been made available (Haunschild & Bornmann, 2021a). We thank the anonymous referees for helpful comments that improved our manuscript.

**Conflicts of Interest / Competing Interests**

LB is member of the Price Medal Laureates Board of this journal, and RH is member of the Distinguished Reviewers Board of this journal.